# Van der Waals waveguide quantum electrodynamics probed by infrared nano-photoluminescence

S. L. Moore[1*], H. Y. Lee[2,3], N. Rivera[4], Y. Karube[5], M. Ziffer[3], E. S. Yanev[3], T. P. Darlington[3], A. J. Sternbach[6], M. A. Holbrook[1,3], J. Pack[1], X. Xu[7], C. R. Dean[1], J. S. Owen[5], P. J. Schuck[3], M. Delor[5], X.Y. Zhu[5], J. Hone[3], D. N. Basov[1]

[1]Department of Physics, Columbia University. [2]Department of Materials Science and Nanoengineering, Rice University. [3]Department of Mechanical Engineering, Columbia University. [4]Department of Physics, Harvard University. [5]Department of Chemistry, Columbia University [6]Department of Physics, University of Maryland. [7]Department of Physics, Department of Materials Science and Engineering, University of Washington.
*Corresponding author: slm2236@columbia.edu

**Abstract**
Atomically layered van der Waals (vdW) materials exhibit remarkable properties, including highly-confined infrared waveguide modes and the capacity for infrared emission in the monolayer limit. Here, we engineered structures that leverage both of these nano-optical functionalities. Specifically, we encased a photoluminescing atomic sheet of $MoTe_2$ within two bulk crystals of $WSe_2$, forming a vdW waveguide for the embedded light-emitting monolayer. The modified electromagnetic environment offered by the $WSe_2$ waveguide alters $MoTe_2$ spontaneous emission, a phenomenon we directly image with our interferometric nano-photoluminescence technique.  We captured spatially-oscillating nanoscale patterns prompted by spontaneous emission from $MoTe_2$ into waveguide modes of $WSe_2$ slabs. We quantify the resulting Purcell-enhanced emission rate within the framework of a waveguide quantum electrodynamics (QED) model, relating the $MoTe_2$ spontaneous emission rate to the measured waveguide dispersion.  Our work marks a significant advance in the implementation of all-vdW QED waveguides.

**Introduction**

Spontaneous emission of atoms, molecules or solids can be altered by precisely engineering the electromagnetic environment of these emitters with cavities or waveguides. The many roles of the electromagnetic environment in the emission are often categorized by the Purcell effect[1–4] . The QED description of electromagnetic cavities established that the efficacy of the Purcell effect is governed by the ratio $Q/V_m$, where $Q$ is the cavity quality factor, and $V_m$ is its mode volume.  Accordingly, the extreme subwavelength confinement (low $V_m$) enabled, for example, by nano-cavity plasmons produces some of the largest reported Purcell enhancements[5–10].  However, plasmonically-controlled emission inevitably carries significant loss (i.e. low $Q$) inherent to plasmons in metals. Dielectric cavities and waveguides circumvent the loss problem and achieve much larger values of $Q$, enabling confined light to propagate over considerable distances. However, sophisticated metastructure designs[11–15] are typically required to achieve notable Purcell enhancements in all-dielectric waveguides. Remarkably, high refractive index van der Waals (vdW) materials seamlessly integrate both superior confinement and low loss[16–19], effectively serving as natural cavities or waveguides in the infrared frequency range (0.05 – 1 eV). Nevertheless, the ability of natural vdW waveguides to promote Purcell enhanced photoluminescence (PL), especially at technologically important infrared frequencies, remains largely unstudied because of technical limitations associated with probing both spatially and spectrally resolved emission at the nano-scale.

Here we report on waveguide-assisted Purcell effects in the infrared range, harnessing transition metal dichalcogenides (TMDs) as both emitters and waveguides.  Because of their high refractive indices, TMDs enable total internally-reflected light to propagate within waveguides with wavelengths 3-5 times smaller than free space light[18,20].  The dispersion of waveguide modes in TMDs is governed by the slab thickness and can be adjusted in the exfoliation process. Upon thinning down to the monolayer limit, many TMDs, including 2H-$MoTe_2$, exhibit a direct band gap (Fig. 1a,b) and emit PL with high quantum efficiency[21]. Using the dry-transfer stacking technique (methods), we utilized the two planar slabs of $WSe_2$ with judiciously chosen thicknesses to encase the monolayer of MoTe$_2$ (Fig.1c). This procedure enables deterministic control of the emitting layer vertical placement within the waveguide, thereby optimizing the Purcell enhancement. Losses in TMD materials are relatively low when probed below the optical gap (commonly in the 1 – 2 eV range). However, experimental methods for spatially resolving PL from buried emitting layers in the infrared range presents a technical challenge. Thus, direct imaging of Purcell-enhanced infrared PL reported in Fig.1 remained insofar elusive.

We overcome challenges of nano-PL spectroscopy and imaging of waveguided structures by utilizing experiments outlined in Fig.1c, facilitated by a recently developed antenna-based imaging tool that combines nano-PL and nano-scattering modalities[22]. These antennas find practical implementation through the sharp tips of atomic force microscopes. To date, antenna-based nano-PL[6,22–26] data have been collected from emitters directly underneath the tip, inducing local near-field PL enhancement. Here, aided by the high-index/low-absorption vdW waveguides, we utilize nano-PL to spatially and spectrally resolve emission from a uniformly-excited buried-layer into confined waveguide modes of the surrounding TMD slab. In our data, the waveguiding manifests as a dispersing interference pattern, implying coherent light propagation within the examined sample volume. We augmented the nano-PL data with collocated nano-scattering dispersion mapping and analytical Purcell-factor modeling. The totality of the observations and modeling establish the first nano-scale imaging of Purcell-enhanced emission into coherently propagating waveguide modes. Our experiments set a precedent for the imaging and Purcell analysis of emission in van der Waals waveguide-QED platforms.

**Hyperspectral Nano-Photoluminescence Imaging**

Monolayer 2H-MoTe$_2$ is chosen as a relatively narrow bandgap emitter, and the larger bandgap material WSe$_2$ is selected for the receiver waveguide. Fig. 1a illustrates that type-I band alignment in our waveguide devices promotes charge transfer from WSe$_2$ to MoTe$_2$, potentially yielding enhanced emission intensity[27,28]. The dipoles activated in MoTe$_2$ by the excitation laser with energy $\Omega = 1.65$eV emit ~0.8-1.05 eV light into the WSe$_2$ waveguide. Figure 1b displays the room-temperature imaginary (dissipative) part of the dielectric function, $\epsilon_2(\omega)$, for bulk WSe$_2$ and monolayer MoTe$_2$ (methods, Supplementary Section 1). The data confirm the MoTe$_2$ exciton at $E_{1s} = 1.09$ eV is well below the $E_{1s} = 1.59$ eV exciton resonance in WSe$_2$.

We display our antenna-based nano-optical setup in Fig.1c, with further details provided in the methods and Supplementary Section 2. The WSe$_2$/MoTe$_2$/WSe$_2$ structures rest on template-stripped gold (methods), which improves both the PL excitation and the collection of waveguided emission. The excitation laser illuminates the entire sample in the far field, initiating emission from the MoTe$_2$ monolayer. As illustrated in Fig. 1c, the MoTe$_2$ may be expected to emit directly into the WSe$_2$ waveguide modes. This waveguided emission then propagates over macroscopic distances in the form of well-defined waveguide modes and eventually reaches the tip or top WSe$_2$ edge. Both the tip and the WSe$_2$ edge outcouple waveguide modes, transforming them into free-space light that is detected by an InGaAs CCD/spectrometer. Each emitting MoTe$_2$ dipole utilizes both tip- and edge-enabled outcoupling channels, producing interference at the detector set by the emission frequency $\omega$ and tip-edge distance $x$. By raster scanning the tip, one maps out the corresponding phase-difference $\Delta\phi(x,\omega)$ at each frequency $\omega$, generating a dispersing interference pattern.

The hyperspectral nano-PL line-profile collected at room-temperature (Fig. 1d) reveals dispersing subwavelength oscillations, a direct measure of $\Delta\phi(x,\omega)$. We plot the differential intensity $I(x,\omega)$, defined as the difference between the in-contact PL intensity at tip position $x$ and the PL signal with the sample retracted from the tip by 150 nm. The $I(x,\omega)$ data uncover dispersing oscillating fringes (Fig.1d) that are most prominent near the edge of the top WSe$_2$ layer at $x = 0$. Oscillations abruptly cease to the left of the edge but extend into the interior of the WSe$_2$ waveguide for several microns. An un-normalized far-field PL spectrum from the same dataset is superimposed along the y-axis, showcasing broadband emission centered at

1.04 eV. We assert that the oscillations observed in Fig.1d likely originate from MoTe$_2$ emission into distinct propagating modes supported by the WSe$_2$ waveguide and will justify this assertion with data reported in Fig. 2.

To verify the hypothesis of MoTe$_2$ emission into WSe$_2$ waveguide modes, we characterized these propagating modes using infrared nano-scattering experiments collocated with nano-PL results. Nano-infrared scattering, or nano-IR, is an established methodology for mapping dispersive confined modes in vdW heterostructures (see methods for details)[18,29,30]. In Figure 2a, we plot the nano-IR images collected at multiple laser energies within the MoTe$_2$ emission bandwidth. We observe a series of oscillating and dispersive profiles in our nano-IR data. For quantitative comparison, we apply Fourier analysis to both the nano-IR and nano-PL datasets. To decode these dispersions, we identify the set of theoretically possible TE and TM modes, as exemplified in Fig. 2b, for a WSe$_2$ waveguide with the total thickness of $d = 98$ nm (see Supplementary Section 3 for modeling details). Each mode $\ell = 0,1,2\cdots$ in Fig. 2 is fully defined by a complex wavevector $q_{\sigma\ell} \equiv q_{1\sigma\ell} + iq_{2\sigma\ell}$, where the index $\sigma$ = TE, TM stands for the polarization. The associated mode wavelength $\lambda_q \equiv 2\pi/q_{1\sigma\ell}$ and propagation length $L_q \equiv 1/q_{2\sigma\ell}$ can be evaluated knowing the complex wavevectors. As typical for high refractive index waveguides[18], $\lambda_q$ in our structures is about 3 times shorter than the free space wavelength. The TE modes are bounded by the WSe$_2$ light-cone (Fig. 2b, right-most black dashed line), while the TM modes are bounded by the WSe$_2$/Au surface plasmon polariton (SPP, green dashed line). Non-waveguided "air" modes[31], meanwhile, are bounded below the vacuum light line (gray shaded area). In nano-IR experiments, a given mode with indices $\sigma, \ell$ is known to produce two distinct spatial periodicities $\lambda$ and $\lambda/2$. The $\lambda$ oscillations arise from interference between photons outcoupled by the edge and by the tip, originally illustrated by red arrows in Fig. 1c. On the contrary, $\lambda/2$ patterns originate from modes that are both launched and outcoupled by the tip, completing a roundtrip between the tip and the edge. In Supplementary Section 3, we relate these "$\lambda$-" and "$\lambda/2$-" modes to the theoretical dispersion $(\omega, q_{1\sigma\ell})$ of Fig. 2b, allowing us to overlay these predictions onto the Fast-Fourier-Transformed (FFT) nano-IR and nano-PL spatial profiles (Figs. 2c and 2d, respectively). Both nano-scattering and nano-PL data reveal nearly identical dispersions. When the waveguide thicknesses are altered, both nano-IR and nano-PL dispersions change accordingly (Supplementary Section 4, Fig. S4). In Supplementary Section 4, we also comment on the relative amplitudes of $\lambda$ and $\lambda/2$ modes observed between the nano-IR and nano-PL channels. We conclude that in our multilayer structures, the MoTe$_2$ monolayer emits into the same waveguide modes of WSe$_2$ that govern nano-IR scattering experiments.

**Nano-PL modeling**

We now outline the key aspects of our nano-PL experiments that are important to set up a proper model. Nano-PL signals reported in Figs. 1-2 all occur in the regime of large separation between MoTe$_2$ and our antenna nano-probe. This is in contrast with more common antenna-based nano-PL experiments relying on near-field interactions between the tip and emitter[25,32]. Here, the use of confined propagating modes facilitates sub-wavelength readout of buried (30-250 nm deep) emitters. Thus, the antenna of our near field apparatus does not couple to the monolayer in the near-field. Instead, the antenna visualizes light emitted into waveguide modes undergoing total internal reflections, which are outcoupled by the tip as the waveguided radiation reaches the top surface. In Supplementary Section 5, we present additional data and simulations supporting this "passive" readout scenario, including the invariance of the observed interference pattern with respect to excitation polarization. We are therefore operating in an unexplored regime for nano-PL detection: a uniformly excited distribution of dipoles, each emitting into the waveguide medium.

We derived an expression for the dispersing interference patterns within the proposed passive tip-readout scenario; this expression expands upon the qualitative discussion of Fig. 1c and quantitatively accounts for the data in Fig. 1d. We assume that excited dipoles are uniformly distributed inside the semi-infinite monolayer. Then, it suffices to consider a one-dimensional line of emitting dipoles oriented perpendicular to the edge of our structure. The waveguide mode originating from an emitter element at coordinate $x_d$ travels to the edge of the crystals at $x = 0$ and also towards the tip located at $x$ (Fig. 1c). Both the edge and tip can outcouple a set of $m$ waveguide modes with complex wavevector $q_m(\omega) = q_{1m}(\omega) + iq_{2m}(\omega)$ into free space photons, indicated with arrows in Fig. 1c. Here, we consider the interference situation for round-trip $\lambda/2$ modes, where modes reflect from the edge back to the tip. Let $\Delta x \equiv x_d + x - |x - x_d|$ denote the path difference between the outcoupled "beams", and let $\Sigma x \equiv x_d + x + |x - x_d|$ be the net propagation distance. We also account for geometric $\sim 1/\sqrt{x}$ attenuation of the propagation within the slab[33]. The interference intensity may then be written $i_m(x, x_d, \omega) \propto e^{-q_{2m}\Sigma x} \cos(q_{1m}\Delta x)/\sqrt{(x_d + x)|x - x_d|}$. Since $\Sigma x = \{2x, 2x_d\}$ and $\Delta x = \{2x_d, 2x\}$, we expect the $\lambda/2$ interference to oscillate with the wavelength $\lambda_m = \pi/q_{1m}$ and decay over a length $L_m = 1/(2q_{2m})$. Note that when interference signals are instead collected from non-guided radiation ($q_{1m} \leq \omega/c$), attenuation will be dominated by geometric decay. The net mode intensity at the detector $I_m^{(\lambda/2)}(x, \omega)$, integrated over all excited dipoles, can be written in the form (See Supplementary Section 5 for further details, including for edge-outcoupled $\lambda$ modes):

$$I_m^{(\lambda/2)}(x, \omega) = A_m[K_0(2q_{2m}x)\cos(2q_{1m}x) + (\pi/2)e^{-2q_{2m}x}J_0(2q_{1m}x)] \quad (1),$$

where $K_0(\cdots)$ and $J_0(\cdots)$ are Bessel functions. The nano-PL data (e.g. Fig. 1d) reveal a series of modes $I(x, \omega) = \sum_m I_m^{(\lambda/2)}(x, \omega)$. Because $I(x, \omega)$ is directly accessible in our experiments, coefficients $A_m, q_{1m}, q_{2m}$ can be extracted by fitting to Eq. 1. Fig. 3a displays a representative fit to an $\omega = 1.03$ eV nano-PL profile for a single $\lambda/2$ mode. In this structure, the MoTe$_2$ monolayer was placed ~1 $\mu$m to the right from the top WSe$_2$ edge; therefore, the fit is constrained to $x > 1.5$ $\mu$m. Slowly varying (extrinsic) backgrounds are suppressed by applying the Laplacian $\partial_x^2 I(x, \omega)$ to experimental data. The fitting parameters $\lambda_m$ and $L_m$ are reported in Fig. 3a. The spectral dependence of the amplitude parameter $A_m$ is examined in Supplementary Section 5. The fit is adequate and quantitatively verifies that emitted light is indeed guided over long propagation distances, as expected in the regime where $q_{2m} \ll q_{1m}$. In Figs. 3b and 3c, we repeat this fitting for all emission energies within the PL bandwidth. Fig. 3b affirms that the wavevector $q_{1m}(\omega)$ agrees well with the predicted TE$_0$ dispersion, a finding that remains unchanged for two different excitation energies $\Omega = 1.72$ eV, 1.63 eV. We may conclude that Eq. 1 reliably captures gross features intrinsic to the waveguided emission.

In Supplementary Section 8, we examine whether PL from our monolayer MoTe$_2$ emitter can be initiated via waveguide modes of WSe$_2$ slabs launched by the excitation laser. Our modeling (Supplementary Section 5) suggests that as the excitation energy $\Omega$ decreases below the WSe$_2$ band edge, undamped waveguide modes emerge with spatial periodicity $2\pi/q_1(\Omega)$. Consequently, the MoTe$_2$ excitation becomes spatially modulated, producing hyperspectral patterns that disperse with excitation energy $\Omega$ rather than emission energy $\omega$. This excitation-modulation effect is most pronounced when the MoTe$_2$ layer is positioned near the gold surface, where the TM-polarized waveguide mode is most intense. Supplementary Section 8 therefore suggests that Eq. 1 is most reliable for excitation above the WSe$_2$ gap and for emitter placements inside the waveguide.

Finally, our analysis of propagation losses in Fig. 3c sets a lower bound on the coherence time $\tau_c$, the time required for a single emitter to lose phase coherence. If emission propagates in a vdW waveguide with a group velocity $v_g$, it will reach a distance $x_c = v_g \tau_c$ before decohering. Supplementary Section 5 generalizes Eq. 1 to the scenario of an emitter with a finite coherence length $x_c$. Since propagation lengths $L_m = 1/q_{2m}$ measured with an external laser (nano-IR, triangles of Fig. 3c) are similar to those measured with nano-PL (squares, circles, Fig. 3c), we conclude that ohmic losses in our waveguides, quantified by $q_{2\sigma\ell}$, dominate the observed nano-PL decay. Therefore, the single-emitter coherence time satisfies $\tau_c > L_m/v_g = 150$ fs (taking $L_m = 9$ $\mu$m from Fig. 3a and $v_g = d\omega/dq_1 = 0.2c$ from Fig. 3b). This bound on $\tau_c$ is longer than the time scale $T_2^*$ set by the far-field, inhomogeneously-broadened PL linewidth (y-axis of Fig. 1d), $T_2^* \sim 40$ fs. Previously, the plausibility of room temperature coherence has been demonstrated in monolayer TMDs embedded in microcavities[34]. We may conclude the observation of coherent propagating emission within the scan range in our slow light waveguides.

**Purcell Analysis of Confined Photoluminescence Propagation**

In the previous section, we established that nano-PL data report the mode dispersion accounted by Eq. 1. Here we calculate a more fundamental QED quantity, the Purcell factor[3,35,36], and discuss its relationship with the mode dispersion $q_{\sigma\ell}$. First, we derive an expression for the Purcell factor $F_m(\omega)$, dependent on energy $\omega$ and labeled by mode index $m$. Next, we evaluate the mode-aggregated Purcell factor $F = \sum F_m$ at a series of thicknesses $d$ of our waveguides and compare against far-field lifetime data. With confidence in the model, we use $F_m(\omega)$ to benchmark the performance of our waveguides examined with nano-PL.

The full derivation of the Purcell factor $F$ is presented in Supplementary Section 6. Here, we define $F \equiv \gamma_{rad}(d)/\gamma_0$, the enhancement of the spontaneous emission rate $\gamma_{rad}$ in a waveguide of thickness $d$ compared to the vacuum emission rate $\gamma_0$. We decompose $F$ into mode-specific contributions $F = F_{ng} + \sum F_{\sigma\ell}$, where the label $\sigma$ denotes the polarization, while $\ell$ is the mode index. Non-guided (air mode) emission is also considered and is denoted by $F_{ng}$. We separately derived an equation for the so-called nonradiative Purcell factor, $\Phi \equiv \gamma_{nrad}(d)/\gamma_0$, where $\gamma_{nrad}$ is the nonradiative recombination rate into Ohmic currents of surrounding media. Together, $\Phi_{\sigma\ell}$ and $F_{ng}$ reduce the guided-mode spontaneous emission intensity by the factor[3,37] $\beta \equiv \sum_{\sigma\ell} F_{\sigma\ell}/(\Phi + F)$. Exact expressions can be found in supplementary equations S6.42-S6.43. Approximate expressions for $F_{\sigma\ell}$ and $\Phi_{\sigma\ell}$ in the regime of low waveguide dissipation ($q_{2\sigma\ell} \ll q_{1\sigma\ell}$) obey:

$$F_{\sigma\ell} \approx \frac{3\pi}{4} n_{g\sigma\ell} \left(\frac{cq_{1\sigma\ell}}{\omega}\right) \left(\frac{c}{\omega L_{\sigma\ell}}\right) |e_{\perp\sigma\ell}(z)|^2 \quad (2)$$

$$\Phi_{\sigma\ell} \approx \frac{3\pi}{8} n_{g\sigma\ell}^2 \left(\frac{\Lambda_{\sigma\ell}}{L_{\sigma\ell}}\right) \left(\frac{q_{1\sigma\ell}}{q_{2\sigma\ell}}\right) \left(\frac{c}{\omega L_{\sigma\ell}}\right) |e_{\perp\sigma\ell}(z)|^2 \quad (3)$$

(Supplementary Section 6 demonstrates close agreement between Eqs. 2, 3, and exact expressions S6.42, and S6.43). Here, $e_{\perp\sigma\ell}(z)$ is the in-plane electric field component of the waveguide eigenmode $\mathbf{e}_{\sigma\ell}(z) \equiv e_{\perp\sigma\ell}(z)\hat{\rho} + e_{z\sigma\ell}(z)\hat{z}$. Next, $L_{\sigma\ell} \equiv \sum_{k=\perp,z} \int_{-\infty}^{\infty} dz' \epsilon_{1k}(z')|e_{k\sigma\ell}(z')|^2$ is the out-of-plane mode confinement, where $\epsilon_k(z) = \epsilon_{1k}(z) + i\epsilon_{2k}(z)$ is the $k$-th principal component of the dielectric tensor at height $z$ above the substrate. Finally, $n_{g\sigma\ell} \equiv c(dq_{1\sigma\ell}/d\omega)$ is the group index, and $\Lambda_{\sigma\ell} \equiv \sum_{k=\perp,z} \int_{-\infty}^{\infty} dz' \epsilon_{2k}(z')|e_{k\sigma\ell}(z')|^2$ is the nonradiative loss coefficient.

Several points can be drawn from Eqs. 2 and 3. First, in the limits $|e_{\perp\sigma\ell}(z)|^2 \sim 1$ and $L_{\sigma\ell} \sim d/n_{\text{eff}}$, the Purcell factor—a parameter commonly extracted from far-field lifetime experiments—can readily be estimated from the measured waveguide dispersion $q_{1\sigma\ell}(\omega)$ knowing the thickness $d$ and effective index $n_{\text{eff}} \sim \sqrt{\epsilon_\perp}$. Second, in the presence of finite loss ($\Lambda_{\sigma\ell} > 0$), an increase in $q_{1\sigma\ell}$ enhances both radiative and nonradiative recombination. Third, the emitter vertical position $z$ must be carefully chosen to maximize $e_{\perp\sigma\ell}(z)$, a straightforward optimization in our all-vdW waveguide-QED platform. (In all waveguides presented here, the MoTe$_2$ is placed at $z \approx 0.55d$. See Supplementary Section 8 for nano-PL data with different emitter positions). Note that the apparent $1/q_2$ divergence in Eq. 3 is remedied by considering instead the limit $\epsilon_2(z) \to 0$, where $\Lambda_{\sigma\ell}$ scales linearly with $\epsilon_2(z)$, and $q_{2\sigma\ell}$ scales sub-linearly. Thus, the ingredients for computing these separate contributions to the Purcell factor rely on inputs readily determined from our experiments. This correspondence highlights the utility of nano-PL imaging for waveguide-QED applications.

As described in Figures 4a-b, the quantity $\gamma(d)/\gamma(\infty) = [F(d) + \Phi(d)]/n_{\text{eff}}$ exhibits gross agreement with far-field time-resolved PL data (TRPL, see methods), where $\gamma(d) \equiv \gamma_{rad}(d) + \gamma_{nrad}(d)$. A TRPL trace in Fig. 4a displays raw PL counts at time $t$ after 5 ps-long pulsed excitation with energy centered at $\Omega = 1.77$ eV. We operated at reduced power to avoid sample damage and Auger (many-body) nonradiative recombination. Emission intensity is then expected to follow a mono-exponential decay $\propto e^{-\gamma(d)t}$, yet the data generally showcase at least two separate decay rates. We propose, as illustrated by the straight dashed lines, that the fast decay time (100 ps) corresponds to the exciton (X) and the secondary time (500 ps) to the trion (X*). This assignment is consistent with previously published data for both isolated TMD monolayers and similar MoTe$_2$/WSe$_2$ heterostructures[27,38]. In our nano-PL data, we probe waveguiding below both the exciton and trion resonances; low temperature measurements will be required to spectrally separate these two contributions. In Supplementary Section 7, we present raw TRPL traces for structures with different waveguide thicknesses, revealing a systematic increase in these decay rates for thin waveguides. Fig. 4b presents the normalized decay rates derived from TRPL for structures with several thicknesses. Experimentally, $\gamma(\infty)$ corresponds to the decay rate from the thickest waveguide fabricated in this work, $\gamma(528\text{ nm})$. The error bars in the data points stem from the standard deviation in the measurements carried out at three different locations in the interior of MoTe$_2$.

The agreement between the TRPL data and the theoretical model $\gamma(d)/\gamma(\infty) = [F(d) + \Phi(d)]/n_{\text{eff}}$ in Fig. 4b establishes confidence in using Eqs. 2-3 or S6.42-43 to describe mode-specific trends of the Purcell factor. In Fig. 4c, we evaluate the mode-specific Purcell factor versus waveguide thickness $d$. On the right-hand vertical axes, we divide $F_{\sigma\ell}$ by the effective WSe$_2$ index $n_{\text{eff}}$. The quantity $\gamma_{\text{rad}}(d)/(n_{\text{eff}}\gamma_0)$ signifies the Purcell-enhanced emission rate, $\gamma_{rad}(d)$, relative to the bulk emission rate $\gamma(\infty) = n_{\text{eff}}\gamma_0$. We first consider the behavior of the mode-aggregated factors $F$ (solid black line). We fix the energy to $\omega = 1.03$ eV (Supplementary Section 6 details the Purcell factor energy dependence) and the emitter vertical position to $z = 0.55d$ (near-midpoint of the waveguide). The black markers on the top axis reference the waveguide thicknesses assembled in this work. Fig. 4c demonstrates that the optical confinement from thin WSe$_2$ waveguides accelerates the MoTe$_2$ spontaneous emission rate up to 1.8 times faster compared to that of bulk ($d \to \infty$) WSe$_2$ slabs. Likewise, the Purcell factor relative to free space reaches 6. Thanks to the optimal emitter placement $z$, the non-guided radiation rate (blue curve) is suppressed at most thicknesses. Next, we consider the subset of guided modes that contribute strongest to the Purcell factor. At low thicknesses, $50 < d < 200$ nm, we observe prominent TE$_0$ waveguiding in nano-PL (e.g. Fig. 2d), consistent with the sizable TE$_0$ Purcell factor in Fig. 4c. At larger thicknesses, higher-order modes produce significant contributions to the Purcell factor, consistent with our extended nano-PL dataset in Supplementary Section 4. The modeling of Fig. 4c suggests favorable Purcell enhancements for the modes captured with our nano-PL technique.

According to Eqs. 2-3, an increased Purcell enhancement often comes at the cost of nonradiative quenching $\Phi(d)$, reducing the radiative yield $\beta$. In Supplementary Section 6, we plot $\Phi_{\sigma\ell}(d)$, finding that maxima in $F(d)$ and $\Phi(d)$ coincide. Notably, the TM$_0$ mode makes the largest contribution to the nonradiative Purcell factor. As described previously, TM$_0$ momenta are bounded below the surface plasmon at the WSe$_2$/Au interface (green dashed line of Fig. 2b), suggesting that quenching may arise from plasmonic coupling to the gold substrate. Supplementary Section 7 argues that quenching from waveguide dissipation also contributes to $\Phi(d)$. In Fig. 4d, we plot the mode-specific radiative yield $\beta_{\sigma\ell} \equiv F_{\sigma\ell}/(F + \Phi)$. We find that each $\beta_{\sigma\ell}$ is optimized in the intermediate thickness range $98 < d < 400$ nm, where nonradiative quenching is low. The mode-aggregated radiative yield (the "$\beta$-factor") ranges from 0.3 to 0.7 for the waveguide thicknesses assembled in this work. In Supplementary Section 9, we show that these values exceed current state-of-the-art vdW waveguides. There, we discuss ways to improve the $\beta$-factor, e.g. low temperature and h-BN spacers. Owing to the minimal non-guided radiation factor (blue curve of Fig. 4c), we anticipate $\beta \to 0.95$ upon suppressing $\Phi(d)$. Such high $\beta$-factors will be critical for future nano-optical investigations of collective coherence between emitters.

## Summary and outlook

In summary, we have used nano-PL to visualize a celebrated phenomenon of waveguide-QED: Purcell-enhanced spontaneous emission into high-index waveguide modes.

We have extended the nano-PL technique to the lowest emission energy (0.8-1.1 eV) measured to date. While type-II band-aligned heterostructures result in PL quenching, the type-I band alignment here enables bright dispersing emission oscillations that we visualized with hyperspectral nano-PL imaging. We observe prominent subdiffractional waveguiding of the $MoTe_2$ emission, despite the large tip-emitter distance. The corresponding interference patterns imply coherent propagation over the spatial extent of our experiments. Our analytical Purcell analysis confirms that the electromagnetic environment of the $WSe_2$ slabs significantly alters spontaneous emission from the embedded $MoTe_2$ monolayer. Within this model, we identified the range of thicknesses to optimize waveguided radiation efficiency, $\beta \sim 0.8$, such that $TE_0$ radiation dominates and $TM_0$ non-radiative channels are subdued. The maximum Purcell factor of 6 is comparable to state-of-the-art waveguides (Supplementary Section 9) and achieved without complex lithographic patterning. However, to mitigate geometric losses and further enhance the Purcell factor, future experiments could incorporate lithographic patterning to impose unidirectional propagation on the emission.

Our work underscores the promise of vdW materials and interfaces for future waveguide-QED research. There remains an unexplored catalogue of ultra-confined propagating modes[39] in vdW materials, and the presented Purcell model (Eq. 2) provides analytical predictions for optimizing their couplings to embedded emitters. Highly confined hyperbolic polaritons, including those found in the mid-infrared range (h-BN[40]) or the near-infrared range (CrSBr[31]), are particularly exciting for realizing unparalleled Purcell and $\beta$ factors. In CrSBr, the high degree of anisotropy could enable unidirectional waveguiding without lithographic patterning. While the Purcell effect is well-understood within the framework of QED, it can also be interpreted as a classical interference phenomenon[41]. An end goal is therefore to exploit the propagating (extended) nature of confined modes in vdW waveguides to facilitate distinct quantum-optical phenomena, including single-photon emission, collective interactions[42,43] and possibly superradiance[35,44]. In vdW materials, these low-temperature phenomena may be tunable using strain[45] or moiré potentials[46]. Embedding emitters into polaritonic vdW waveguides could unlock novel regimes of waveguide QED.

## Acknowledgments

The authors thank A. Asenjo-Garcia for valuable discussion. Research on vdW waveguides at Columbia is supported as part of Programmable Quantum Materials, an Energy Frontier Research Center funded by the U.S. Department of Energy (DOE), Office of Science, Basic Energy Sciences (BES), under award DE-SC0019443. Development of nano-optical imaging is supported by DoE-BES DE-SC0018426.

## Author Contributions

S. L. Moore wrote the manuscript, performed the experiments, analyzed the data, and fabricated the heterostructures. H. Y. Lee exfoliated the $MoTe_2$ monolayers and guided the project. N. Rivera provided theoretical support for the data analysis. Y. Karube performed the TRPL measurements. M. Ziffer guided the TRPL measurements. E. S. Yanev fabricated the template-stripped gold substrates. T. P. Darlington provided expertise for nano-PL measurements. A. J. Sternbach provided guidance for the data analysis. M. Holbrook synthesized the $WSe_2$ and $MoTe_2$ crystal. J. Pack exfoliated the $WSe_2$ crystal. X. Xu, X. Y. Zhu, J. S. Owen, M. Delor, P. J. Schuck, D. N. Basov, provided guidance for the project. C. R. Dean and J. Hone provided access to lab facilities.

## Competing Interests

The authors declare no competing interests.

## Figure Captions

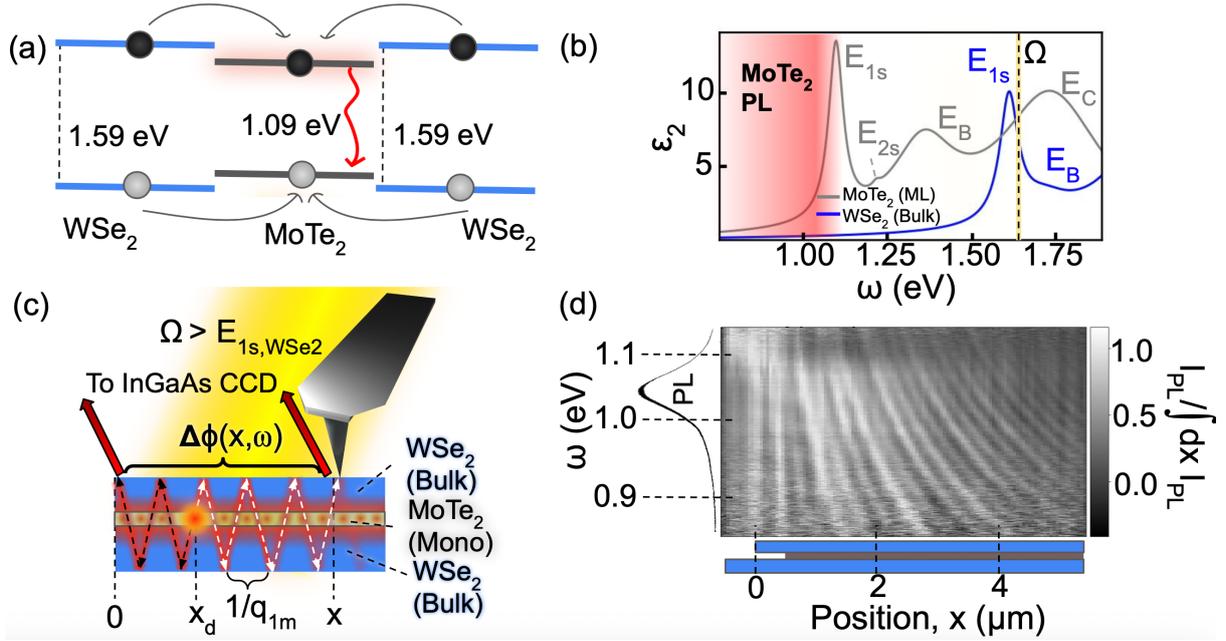

**Figure 1**: Hyperspectral nano-photoluminescence imaging in a WSe$_2$/MoTe$_2$/WSe$_2$ waveguide heterostructure. In (a), we illustrate the band alignment at the WSe$_2$/MoTe$_2$ interface, resulting in charge transfer from WSe$_2$ into MoTe$_2$. Panel (b) plots the imaginary part of the dielectric function $\epsilon_2(\omega)$—obtained by far-field reflectance—for bulk WSe$_2$ and monolayer MoTe$_2$ encapsulated in hBN. The PL energy range is shaded in red, and the excitation laser energy $\Omega$ (yellow dashed line) is chosen to overlap with the WSe$_2$ absorption edge. Various excitonic species are labeled for reference. In (c), we illustrate nano-PL detection of waveguided emission. The excitation laser with energy $\Omega = 1.65$ eV uniformly illuminates the sample. Emission from a specific dipole in the MoTe$_2$ layer (at position $x_d$) is outcoupled both from the tip and the edge of the structure, producing interference in accordance with the phase difference $\Delta\phi$. Panel (d) presents the nano-PL hyperspectral line profile measured near the top WSe$_2$ edge, where $I_{PL}$ is acquired by taking the difference between tip-in-contact and tip-out-of-contact spectra. At each energy $\omega$, the profile is normalized to the spatially-integrated intensity. Note that $x = 0$ corresponds to the top WSe$_2$ edge. The far-field PL spectrum, obtained from the same heterostructure, is overlaid on the y-axis for comparison.

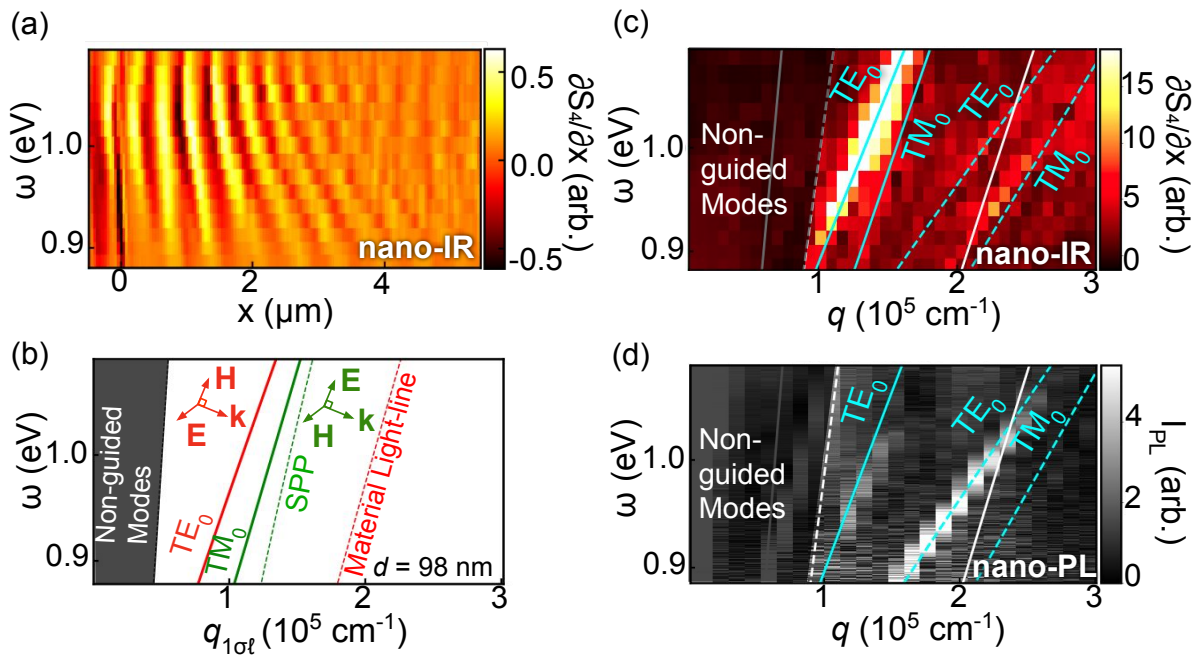

**Figure 2:** nano-PL and nano-IR dispersions for a WSe$_2$/MoTe$_2$/WSe$_2$ waveguide. Panel (a) presents frequency-dependent nano-IR hyperspectral line profiles, $\partial S_4/\partial x$, measured in the same region as Fig. 1d. In (b), we plot the predicted modes for a WSe$_2$/MoTe$_2$/WSe$_2$/Gold heterostructure. The top/bottom WSe$_2$ thicknesses are 46 and 52 nm, respectively. The arrows indicate the polarization of the fields **E**, **H**, both orthogonal to the wavevector **k**. Panels (c) and (d) display Fourier-transformed nano-IR and nano-PL hyperspectral line profiles, respectively. The experimental data (from the same sample as Fig. 1d) are overlaid with the Fig. 2b theoretical TE$_0$, TM$_0$ modes. Solid and dashed lines represent $\lambda$ and $\lambda/2$ modes, respectively. The relation between the theoretical dispersion, Fig. 2b, with the measured $\lambda$ and $\lambda/2$ modes is described in Supplementary Section 3. White lines indicate the vacuum and material light lines.

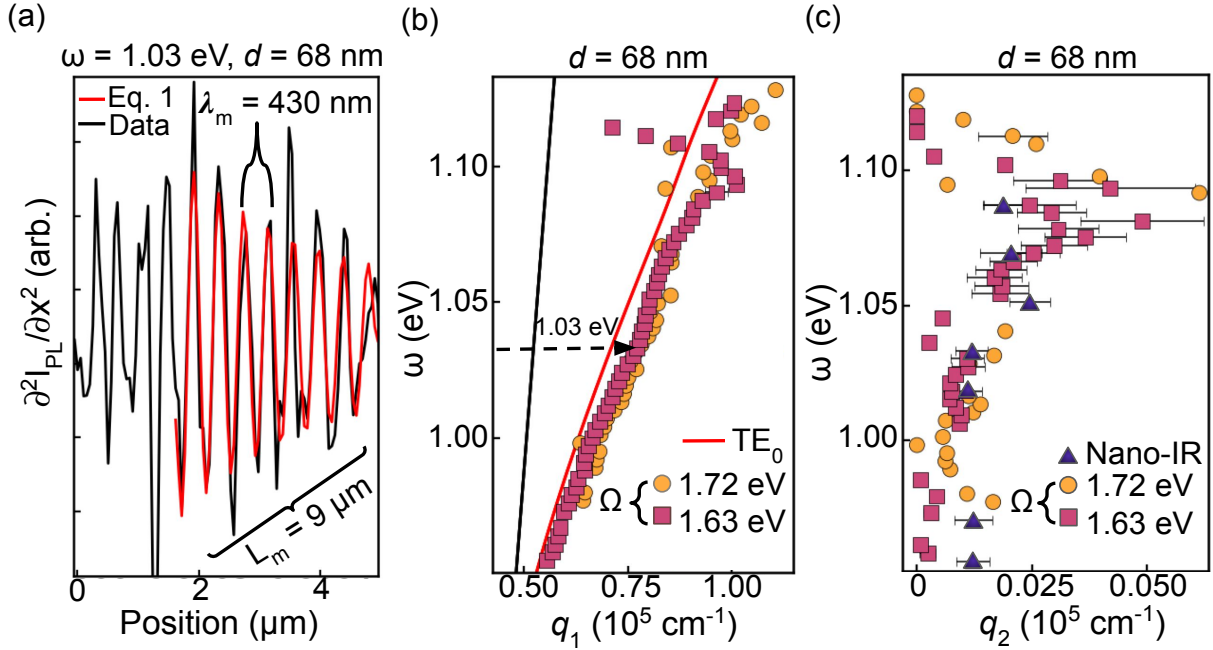

**Figure 3:** Understanding nano-PL profiles. Panel (a) showcases a fit to a 1.03 eV nano-PL line profile for a $\lambda/2$ mode detected in a 68 nm waveguide. The extracted wavelength $\lambda_m$ and propagation length $L_m$ are labeled for reference. In panel (b) we display the nano-PL fitting results $q_{1,m}(\omega)$ at two selected excitation energies $\Omega$. There, the energy $\omega = 1.03$ eV corresponding to the linecut in panel (a) is denoted by a horizontal dashed arrow, and we also include the theoretical TE$_0$ dispersion. The solid black and red lines denote the vacuum light line and theoretical dispersion, respectively. In panel (c), we plot the scattering- and PL-derived imaginary wavevector $q_{2,m=TE0}(\omega)$. Note that all error bars in panels (b)-(c) stem from the confidence intervals in parameters fit to Eq. 1, with the center points defined as values extracted from the fitting. For improved signal-to-noise ratio, the profiles are averaged over three energies within a 2 meV bandwidth.

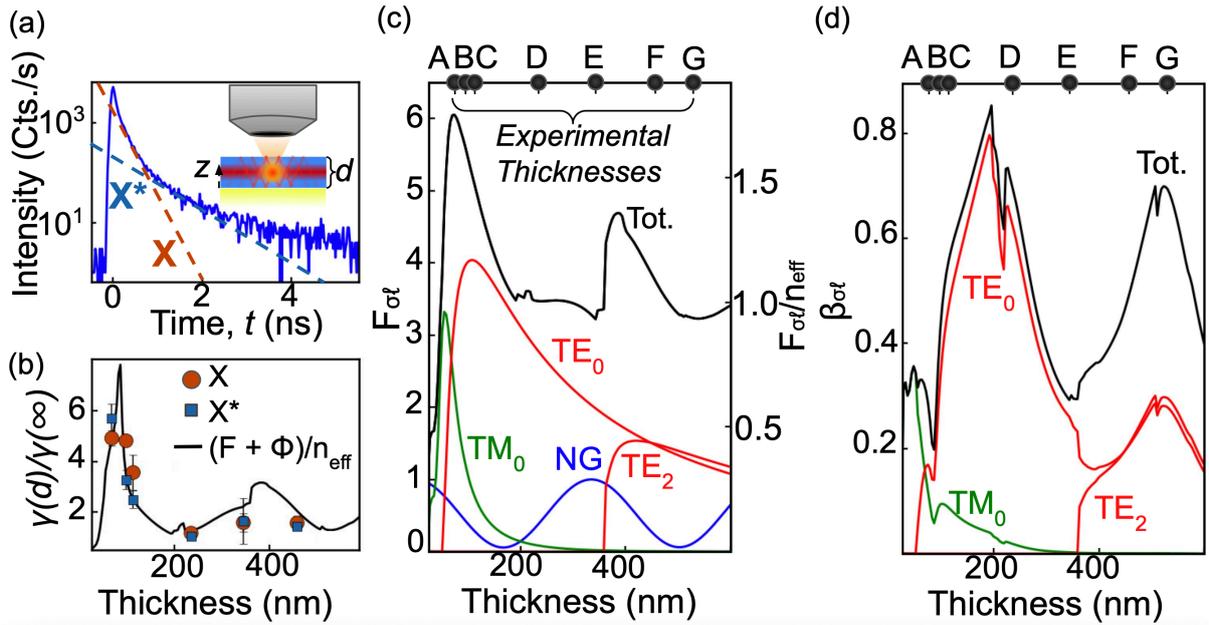

**Figure 4:** Purcell analysis of $WSe_2/MoTe_2/WSe_2$ waveguides. In (a), we show a representative TRPL trace, displaying far-field PL counts at time $t$ after pulsed excitation. The exciton and trion components of the biexponential decay are labeled X and X*, respectively. In panel (b), we display the normalized decay rates $\gamma(d)/\gamma(\infty)$, extracted from TRPL data, for both excitons and trions (data points). The solid curve is given by the theoretical model $[F(d) + \Phi(d)]/n_{\text{eff}}$. The error bars in the data points originate from the standard deviation of the measured lifetime over three separate positions, while the center is determined from the mean lifetime. In panel (c), we display the mode-specific radiative Purcell factors for several selected modes $\sigma, \ell$. The black markers on the top axis refer to the waveguides A-G investigated in this work with thicknesses ranging from 68 nm to 528 nm. The emission energy is fixed to $\omega = 1.03$ eV and the emitter vertical coordinate $z$ is set to the middle of the waveguide ($0.55d$). The solid black line indicates the mode-aggregated Purcell Factor $F$. Panel (d) follows the same labeling conventions as (c), plotting the radiative yield $\beta_{\sigma\ell}$ (solid colored curves) and $\beta$-factor (black curve). Exact expressions S6.42-43 were used to evaluate $F_{\sigma\ell}$ and $\Phi_{\sigma\ell}$ in panels (b-d).

**Methods**

*Pseudo-Heterodyne nano-IR imaging*: A tunable continuous-wave visible – near-infrared laser (M-squared, Inc. 700-2000nm) is sent to a commercial near-field atomic force microscope (Neaspec, Gmbh). The light is split with a 50/50 beam-splitter; half is focused with an NA = 0.4 off-axis parabolic mirror (OAPM) onto a tapping-mode cantilever-based Ag/Au tip (AppNano Inc., 75kHz resonance), and the other half is sent to an oscillating reference mirror. The reference and tip-back-scattered paths interfere at the near-infrared photodiode. The signal is demodulated at harmonics $n = 0 - 5$ of the tapping frequency to produce amplitude $S_n$ and phase $\phi_n$ signals, which in turn can be related to the local reflectivity of the sample[47,48]. Here, $n = 4$ was reported in all datasets for sufficient far-field rejection and signal-to-noise ratio.

*Nano-PL Imaging*: A tunable CW laser follows the same path as in nano-IR imaging to excite the sample. Here, the Ag/Au tip is set to contact mode. Fluorescing light is collected in forward scattering with an NA = 0.4 OAPM, first with the tip in contact, then with the sample retracted by 150nm. The emission is sent directly to a thermoelectrically cooled InGaAs CCD/spectrometer unit (Andor Instruments, Inc., iDus 9/Kymera). Before reaching the spectrometer, the excitation light is blocked with a Thorlabs, Inc. 1100nm OD5 premium long pass filter. To aid in rejection of far-field PL, the filtered light is focused through a 75 micron pinhole. Upon entering the spectrometer, the light is dispersed with a 150 lines/mm grating blazed at 1200nm. Integration times of 6 seconds are typically used, and single line profiles are repeated and averaged up to 10 times for further signal-to-noise improvement. To compensate for spatial drift, rows between consecutive measurements are aligned by tracking the edge position in the AFM topography. Each pixel represents the difference in photoluminescence counts measured in- versus out-of-contact.

*Far-field lifetime measurements*: Photoluminescence lifetime measurements were performed using a custom confocal microscope built around an inverted microscope (Olympus, IX73) fitted with an infinity corrected 40x dry objective (Olympus, LUCPLFLN40x). An NKT Photonics Supercontinuum laser was used to excite the sample with repetition rate 80 MHz, pulse duration 5 ps and wavelength 700 nm. After generating photoluminescence, the laser was filtered out with a 950 nm dichroic longpass mirror (ThorLabs, DMLP950R). Photoluminescence from the sample was coupled to a superconducting nanowire single photon detector (Single Quantum, EOS CS, 1310 nm detector) with an optical fiber (ThorLabs, HI1060). The superconducting nanowire single photon detector was connected to a time correlated single photon counting device (Picoquant, TimeHarp 260). The timing resolution of the instrument is 25 ps.

*Template stripped gold preparation*: Adhesive-free template-stripped gold substrates are prepared by cold welding thin films of gold to each other. This was accomplished by first evaporating gold onto two polished Si/SiO$_2$ wafers. The "bottom" wafer had a 20 nm Au/5 nm Ti coating, while the "top" had a 100 nm of Au no adhesion layer. Both wafers were diced into 1x1 cm chips, which were then cleaned with O$_2$ plasma prior to use. Top and bottom chips were then stacked together with gold layers touching, clamped in a vice, and placed under vacuum. After >10 minutes, the sample was removed, and a razorblade was used to pry the chips apart. This results in delamination of the top gold film onto the bottom gold surface. The roughness of the newly exposed gold surface mimics that of the polished SiO$_2$, which is significantly smoother than an evaporated metal film.

*Sample Preparation*: 2H-MoTe$_2$ monolayers are prepared by exfoliating lab-grown crystals inside a glove box onto O$_2$-plasma treated SiO$_2$/Si substrates. As testament to the high crystal quality of MoTe$_2$, the 2s exciton at $E_{2s} = 1.22$ eV is also observed.
Monolayers of adequate size are then fully encapsulated with the dry transfer stacking technique. A polycarbonate (PC) stamp first picks up a top WSe$_2$ flake, which is then used to pick up the MoTe$_2$ monolayer. For the double-sided waveguide (main text), a second WSe$_2$ flake is picked up over the MoTe$_2$ region. For the single-sided waveguide (Supplementary Section 8), a thin h-BN flake is picked up instead. The transfer slide can then be brought outside of the glovebox without risk of MoTe$_2$ degradation, where the stack is then transferred onto a template-stripped gold substrate. The melted PC is dissolved from the substrate by rinsing in beakers with Chloroform and IPA for ~10 seconds each over several iterations.

**Data Availability**

Example raw nano-PL data and analysis code have been provided in an online data repository[49]. Additional data are available from the corresponding author on reasonable request.

**Code Availability**

Example codes used to process raw nano-PL data and calculate the Purcell factors have been provided in an online data repository[49]. Additional code is available from the corresponding author upon reasonable request.

**Methods References**